\documentclass[12pt]{article}
\usepackage[english,german,french,polish]{babel}
\usepackage[T1]{fontenc}

\begin{document}

\baselineskip 0.75cm
\topmargin -0.6in
\oddsidemargin -0.1in

\let\ni=\noindent

\renewcommand{\thefootnote}{\fnsymbol{footnote}}

\newcommand{\CKM}{Cabibbo--Kobayashi--Maskawa }

\newcommand{\SM}{Standard Model }

\pagestyle {plain}

\setcounter{page}{1}

\pagestyle{empty}

~~~

\begin{flushright}
IFT/00-17
\end{flushright}

{\large\centerline{\bf An alternative: two--mixing texture for three neutrinos}}

{\large\centerline{\bf or three--mixing texture for four neutrinos{\footnote {Supported in part by the Polish KBN--Grant 2 P03B 052 16 (1999--2000).}}}}

\vspace{0.3cm}

{\centerline {\sc Wojciech Kr\'{o}likowski}}

\vspace{0.2cm}

{\centerline {\it Institute of Theoretical Physics, Warsaw University}}

{\centerline {\it Ho\.{z}a 69,~~PL--00--681 Warszawa, ~Poland}}

\vspace{0.3cm}

{\centerline{\bf Abstract}}

\vspace{0.2cm}
The alternative formulated in the title has a chance to be settled, when the existence of the LSND effect is experimentally excluded or confirmed. The first option, much discussed in literature, works in the case of three active neutrinos $ \nu_e\,, \,\nu_\mu\,, \,\nu_\tau$, when among their massive states $ \nu_1\,, \,\nu_2\,, \,\nu_3$ there is no direct mixing between $ \nu_1$ and $ \nu_3 $, and the mass hierarchy $ m^2_1\stackrel{<}{\sim} m^2_2 \ll m^2_3$ holds. This option is  consistent with the observed deficits of solar $ \nu_e$'s  and atmospheric $\nu_\mu $'s, if $ \Delta m^2_{21} \leftrightarrow \Delta m^2_{\rm sol} $ and $ \Delta m^2_{32} \leftrightarrow \Delta m^2_{\rm atm} $. On the other hand, the second option is an extension of the idea of the former to the case of four neutrinos  $ \nu_s\,, \,\nu_e\,, \,\nu_\mu\,, \,\nu_\tau$ (including one sterile neutrino $ \nu_s $), when among their massive states $\nu_0\,, \,\nu_1\,, \,\nu_2\,, \,\nu_3$ there are no direct mixings between $ \nu_0$ and $ \nu_2 $, $ \nu_0 $ and $ \nu_3 $, $ \nu_1$ and $ \nu_3 $, and the mass hierarchy $ m^2_0\stackrel{<}{\sim} m^2_1 \ll m^2_2 \stackrel{<}{\sim} m^2_3 $ is now valid. Such an option, belonging to a class of textures widely discussed in literature, may be consistent with the observed deficits of solar $ \nu_e $'s and atmospheric $ \nu_\mu $'s as well as with the LSND appearance of $ \nu_e $'s in the beam of accelerator $ \nu_\mu$'s, if now $ \Delta m^2_{10} \leftrightarrow \Delta m^2_{\rm sol} $, $ \Delta m^2_{32} \leftrightarrow \Delta m^2_{\rm atm} $ and $ \Delta m^2_{21} \leftrightarrow \Delta m^2_{\rm LSND} $ (however, in the case of solar $\nu_e $'s the role of $ \nu_s $'s in the disappearance of $ \nu_e $'s is recently questioned). In both options, {\it only the close neighbours in the hierarchies of massive neutrinos $\nu_1\,, \,\nu_2\,, \,\nu_3$ and $\nu_0\,, \,\nu_1\,, \,\nu_2\,, \,\nu_3$, respectively, mix directly}. This characteristic feature of the two--mixing texture for three neutrinos or the three--mixing texture for four neutrinos may be somehow physically significant.

\vspace{0.2cm}

\ni PACS numbers: 12.15.Ff , 14.60.Pq , 12.15.Hh .

\vspace{0.6cm}

\ni August 2000

\vfill\eject

~~~
\pagestyle {plain}

\setcounter{page}{1}

\vspace{0.2cm}

\ni {\bf 1. Introduction}

\vspace{0.2cm}

First of all, we would like to emphasize that the alternative formulated in the title of the paper has a chance to be settled, when the existence of the LSND effect [1] is experimentally excluded or confirmed. The first option  of the alternative, much discussed in literature [2], works in the case of three active neutrinos $ \nu_e\,, \,\nu_\mu\,, \,\nu_\tau $, when among their massive states $ \nu_1\,, \,\nu_2\,, \,\nu_3 $ there is no direct mixing between $ \nu_1$ and $ \nu_3 $ [3], and the mass hierarchy $ m^2_1 \stackrel{<}{\sim} m^2_2 \ll m^2_3$ holds. This option is  consistent with the observed deficits of solar $ \nu_e$'s  [4] and atmospheric $\nu_\mu $'s [5], if $ \Delta m^2_{21} \leftrightarrow \Delta m^2_{\rm sol} $ and $ \Delta m^2_{32} \leftrightarrow \Delta m^2_{\rm atm} $. On the other hand, the second option of the alternative is an extension of the idea [3] of the former to the case of four neutrinos  $ \nu_s\,, \,\nu_e\,, \,\nu_\mu\,, \,\nu_\tau$ (including one sterile neutrino $ \nu_s $), when among their massive states $\nu_0\,, \,\nu_1\,, \,\nu_2\,, \,\nu_3$ there are no direct mixings between $ \nu_0$ and $ \nu_2 $, $ \nu_0 $ and $ \nu_3 $, $ \nu_1$ and $ \nu_3 $, and the mass hierarchy $ m^2_0 \stackrel{<}{\sim} m^2_1 \ll m^2_2 \stackrel{<}{\sim} m^2_3 $ is now valid. Such an option, belonging to a class of neutrino textures widely discussed in literature [6], may be consistent with the observed deficits of solar $ \nu_e $'s [4] and atmospheric $ \nu_\mu $'s [5] as well as with the LSND appearance of $ \nu_e $'s in the beam of accelerator $ \nu_\mu$'s [1], if now $ \Delta m^2_{10} \leftrightarrow \Delta m^2_{\rm sol} $, $ \Delta m^2_{32} \leftrightarrow \Delta m^2_{\rm atm} $ and $ \Delta m^2_{21} \leftrightarrow \Delta m^2_{\rm LSND} $ (however, in the case of solar $\nu_e $'s the role of $ \nu_s $'s in the disappearance of $ \nu_e $'s is recently disputed[4,7]). 

In both options, {\it only the close neighbours in the hierarchies of massive neutrinos $\nu_1\,, \,\nu_2\,, \,\nu_3$} [3] {\it and $\nu_0\,, \,\nu_1\,, \,\nu_2\,, \,\nu_3$, respectively, mix directly}. This characteristic feature of the two--mixing texture for three neutrinos or the three--mixing texture for four neutrinos may be somehow physically significant, leading hopefully to a pertinent dynamical model for the neutrino texture. 

\vspace{0.2cm}

\ni {\bf 2. The first option}

\vspace{0.2cm}

If one conjectures that in the generic \CKM$\!\!\!$--type matrix for leptons [8],

%rownanie 1
\begin{equation} 
U =  \left( \begin{array}{ccl} c_{13}c_{12}  & c_{13} s_{12}  & s_{13} e^{-i \delta}  \\ -c_{23} s_{12} - s_{13} s_{23} c_{12}  e^{i \delta} &  \;c_{23} c_{12} - s_{13} s_{23} s_{12}  e^{i \delta}   & c_{13} s_{23}   \\ \;s_{23} s_{12} - s_{13} c_{23} c_{12}  e^{i \delta}  & -s_{23} c_{12} - s_{13} c_{23} s_{12}  e^{i \delta}  & c_{13} c_{23} \end{array} \right) 
\end{equation} 

\ni with $s_{ij} = \sin \theta_{ij} > 0 $ and $c_{ij} = \cos \theta_{ij} \geq 0 $, $ (i\,,\,j = 1,2,3)$, there is practically no direct mixing of massive neutrinos $\nu_1 $ and $\nu_3 $ ({\it i.e.}, $\theta_{13} = 0$), then $ U $ is reduced to the following two--mixing form much discussed previously [2]:

%rownanie 2
\begin{equation} 
U =  \left( \begin{array}{ccc} 1 & 0  & 0  \\ 0 & \; c_{23} & s_{23} \\ 0 & -s_{23} & c_{23} \end{array} \right)  \left( \begin{array}{ccc} \;c_{12} & s_{12}  & 0  \\ -s_{12} & c_{12} & 0 \\ 0 & 0 & 1 \end{array} \right)=  \left( \begin{array}{ccc} c_{12} & s_{12}  & 0  \\ -c_{23} s_{12}  &  \;c_{23} c_{12} & s_{23} \\ \;s_{23} s_{12} & -s_{23} c_{12} & c_{23} \end{array} \right)  \;. 
\end{equation} 

\ni For the two--mixing option (2) the neutrino mixing formula $\nu_\alpha = \sum_i U_{\alpha i} \nu_i $ takes the form

%rownanie 3
\begin{eqnarray}
\nu_{e} & = & c_{12} \nu_1 + s_{12} \nu_2 \;, \nonumber \\
\nu_{\mu} & = & c_{23}(-s_{12}\nu_1 + c_{12} \nu_2) + s_{23} \nu_3 \;, \nonumber \\
\nu_{\tau} & = & -s_{23}(-s_{12}\nu_1 + c_{12} \nu_2) + c_{23} \nu_3 \;, 
\end{eqnarray}

\ni while the inverse neutrino mixing formula $\nu_i = \sum_\alpha U^*_{\alpha i} \nu_\alpha $ gives 

%rownanie 4
\begin{eqnarray}
\nu_{1} & = & c_{12} \nu_e - s_{12}(c_{23} \nu_\mu - s_{23}\nu_\tau) \;, \nonumber \\
\nu_{2} & = & s_{12}\nu_e + c_{12}(c_{23} \nu_\mu - s_{23} \nu_\tau) \;, \nonumber \\
\nu_{3} & = & s_{23}\nu_\mu + c_{23} \nu_\tau \;. 
\end{eqnarray}

\ni Note that Eq. (2) can be presented also in the form $ U = \exp(i \lambda_7 \theta_{23}) \exp(i \lambda_2 \theta_{12}) $, where $\lambda_2 $  and $\lambda_7 $ are two of eight Gell--Mann $3 \times 3$ matrices.

In the representation, where the charged--lepton mass matrix is diagonal (and thus the corresponding diagonalizing matrix --- unit), the lepton mixing matrix $ U = \left( U_{\alpha i} \right) $ $(\alpha = e \,,\, \mu \,,\, \tau\;,\; i = 1,2,3)$ is, at the same time, the diagonalizing matrix for neutrino mass matrix $ M = \left( M_{\alpha \beta} \right) $ $(\alpha\,,\,\beta = e \,,\, \mu \,,\, \tau)$ , $ U^\dagger M U = {\rm diag}(m_1\,,\,m_2\,,\,m_3)$ with  $ m^2_1 \leq m^2_2 \leq m^2_3 $, so that $ M = \left( \sum_i U_{\alpha i} U^*_{\beta i} m_i \right)$. In this case, the orthogonal two--mixing form (2) of $ U $ leads to the real and symmetric

%rownanie 5
\begin{equation} 
\!M\!\! = \!\!  \left(\! \begin{array}{ccc} c_{12}^2 m_1\! +\! s^2_{12} m_2  & (m_2 \!-\! m_1) c_{12} s_{12}c_{23}  & -(m_2 \!-\! m_1) c_{12} s_{12} s_{23} \\ (m_2 \!-\! m_1) c_{12} s_{12}c_{23} &  s^2_{23}m_3 \!+\! c^2_{23}( s^2_{12}m_1 \!+\! c^2_{12} m_2) & ( m_3 \!-\! s^2_{12}m_1 \!-\! c^2_{12}m_2)c_{23} s_{23} \\ -\! (m_2 \!-\! m_1) c_{12} s_{12}s_{23}  & ( m_3 \!-\! s^2_{12}m_1 \!-\! c^2_{12}m_2) c_{23} s_{23} & c^2_{23} m_3 \!+\! s^2_{23}(s^2_{12} m_1 \!+\! c^2_{12} m_2) \end{array}\!\! \right) .
\end{equation} 

\ni Here, as is seen from Eq. (4), the values $ c_{23} = 1/\sqrt{2} = s_{23}$ give maximal mixing of $ \nu_\mu $ and $ \nu_\tau $: $( \nu_\mu \pm  \nu_\tau )/\sqrt{2}$, and then $ c_{12} \simeq 1/\sqrt{2} \simeq s_{12}$ --- a nearly maximal mixing of $ \nu_e $ and $ (\nu_\mu - \nu_\tau)/\sqrt{2} $: approximately $[\nu_e \pm  (\nu_\mu - \nu_\tau)/\sqrt{2}\,]/\sqrt{2} $. 

From the familiar neutrino oscillation formulae

%rownanie 6
\begin{equation} 
P(\nu_\alpha \rightarrow \nu_\beta) = |\langle \nu_\beta| e^{iPL}|
\nu_\alpha\rangle|^2 = \delta_{\alpha \beta} - 4 \sum_{j>i} U^*_{ \beta j} U_{\alpha j}U_{\beta i} U^*_{ \alpha i} \sin^2 x_{j i} \; ,
\end{equation} 

\ni with

%rownanie 7
\begin{equation} 
x_{j i} = 1.27 \frac{\Delta m^2_{j i} L}{E}\; , \; \Delta m^2_{j i} = m^2_j - m^2_i
\end{equation}

\ni  ($\Delta m^2_{j i}$, $L$ and $E$ measured in eV$^2$, km and GeV, respectively) which is valid for $ U^*_{ \beta j} U_{\alpha j}U_{\beta i} U^*_{ \alpha i}$ real (CP violation neglected), one infers in the case of two--mixing option (2) that

%rownanie 8
\begin{eqnarray}
P(\nu_e \rightarrow \nu_e) & = & 1 - (2 c_{12} s_{12})^2  \sin^2 x_{21} \;,
\nonumber \\ 
P( \nu_\mu \rightarrow \nu_\mu) & = & 1 - (2 c_{12} s_{12} c_{23})^2  \sin^2 x_{21} - (2 c_{23} s_{23})^2( s^2_{12}  \sin^2 x_{31} +c^2_{12}  \sin^2 x_{32})    \nonumber \\ 
& \simeq & 1 - (2 c_{23} s_{23})^2  \sin^2 x_{32} \; , \nonumber \\ 
P(\nu_\mu \rightarrow \nu_e) & = &  (2 c_{12} s_{12} c_{23})^2  \sin^2 x_{21} \;,
\end{eqnarray}

\ni where the final step in the second formula is valid for $ x_{32} = x_{\rm atm} = O(1)$, when $ m^2_1 \stackrel{<}{\sim}  m^2_2 \ll m^2_3 $ or equivalently $ \Delta m^2_{21} \ll \Delta m^2_{32} \stackrel{<}{\sim}  \Delta m^2_{31}$ .

The first formula (8) is consistent with the observed deficit of solar $ \nu_e $'s if one applies the smaller--mass or larger--mass vacuum global solution or large--angle MSW global solution or finally LOW global solution [4] with $(2 c_{12} s_{12})^2 \leftrightarrow \sin^2 2\theta_{\rm sol} \sim (0.72 $ or 0.90  or 0.79 or 0.91) and $ \Delta m^2_{21} \leftrightarrow \Delta m^2_{\rm sol} \sim (6.5 \times 10^{-11} $ or $ 4.4 \times 10^{-10} $ or $ 2.7 \times 10^{-5} $ or $1.0 \times 10^{-7}) \, {\rm eV}^2$, respectively. This gives $ c^2_{12} \sim 0.5 + (0.26 $ or 0.16 or 0.23 or 0.15) and $ s^2_{12} \sim 0.5 - (0.26 $ or 0.16  or 0.23 or 0.15), when taking $ c^2_{12} \geq s^2_{12} $.

The second formula (8) describes correctly the observed deficit of atmospheric $ \nu_\mu $'s [5] if $(2 c_{23} s_{23})^2 \leftrightarrow \sin^2 2 \theta_{\rm atm} \sim 1$ and $ \Delta m^2_{32} \leftrightarrow \Delta m^2_{\rm atm} \sim 3.5 \times 10^{-3} \, {\rm eV}^2$, since then $ \Delta m^2_{21} \ll \Delta m^2_{32} \stackrel{<}{\sim} \Delta m^2_{31}$ for $ \Delta m^2_{21}$ determined as in the case of solar $ \nu_e $'s. This implies that $c^2_{23} \sim 0.5 \sim s^2_{23}$  and $ m^2_3 \sim 3.5 \times 10^{-3}\,{\rm eV}^2 $, because $ m^2_{1} \stackrel{<}{\sim} m^2_{2} \ll m^2_{3}$.

Then, the third formula (8) shows that no LSND effect for accelerator $ \nu_\mu $'s [1] should be observed, $ P(\nu_\mu \rightarrow \nu_e) \sim 0 $, since with $ \Delta m^2_{21} \leftrightarrow \Delta m^2_{\rm sol} \sim (10^{-10 }$ or $10^{-10 }$ or $ 10^{-5}$ or $ 10^{-7})\, {\rm eV}^2 \ll \Delta m^2_{\rm LSND}\sim 1 \, {\rm eV}^2 $, one gets $ \sin^2(x_{12})_{\rm LSND} \sim ( 10^{-21}$ or $10^{-19}$ or $10^{-9 } $ or $10^{-14}) \ll \sin^2 x_{\rm LSND} \sim 1 $, while $(2 c_{12} s_{12} c_{23})^2 \sim (0.72 $ or 0.90 or 0.79 or $ 0.91) \times 0.5 >\sin^2 2\theta_{\rm LSND} \sim 10^{-2}$.

In the case of Chooz experiment looking for oscillations of reactor $\bar{ \nu}_e$'s [9], where it happens that $ (x_{32})_{\rm Chooz} = 1.27 \Delta m^2_{32} L_{\rm Chooz}/E_{\rm Chooz} \sim 1 $ for $\Delta m^2_{32} \leftrightarrow \Delta m^2_{\rm atm} $, the first formula (8) leads to $ P(\bar{\nu}_e \rightarrow \bar{\nu}_e)  \sim 1 $, since $ (x_{21})_{\rm Chooz} \ll (x_{32})_{\rm Chooz} \sim 1$ for $\Delta m^2_{21} \leftrightarrow \Delta m^2_{\rm sol} $ ( $ U_{e3} = 0 $ in our case). This is consistent with the negative result of Chooz experiment. We can see, however, that for the actual lepton counterpart of \CKM matrix the entry $ U_{e3}$ may be a potential correction to the two--mixing option (2) ($ |U_{e3}| < 0.2 $ according to the estimation in Chooz experiment).

Further on, we will put $ c_{23} \simeq 1/\sqrt{2} \simeq s_{23}$. Then, from Eq. (5) we infer that approximately

%rownanie 9
\begin{equation} 
M \!=\!  \left( \!\! \begin{array}{ccc} c_{12}^2 m_1 \!+\! s^2_{12} m_2  & (m_2 \!-\! m_1) c_{12} s_{12}/ \sqrt{2} & -(m_2 \!-\! m_1) c_{12} s_{12} / \sqrt{2}   \\ (m_2 \!-\! m_1) c_{12} s_{12}/ \sqrt{2} &  (m_3 \!+\!  s^2_{12}m_1 \!+\! c^2_{12} m_2)/2 & ( m_3 \!-\! s^2_{12}m_1 \!-\! c^2_{12}m_2)/2 \\ -(m_2 \!-\! m_1) c_{12} s_{12} / \sqrt{2} & ( m_3 \!-\! s^2_{12}m_1 \!-\! c^2_{12}m_2) /2 & (m_3 \!+\! s^2_{12} m_1 \!+\! c^2_{12} m_2) / 2 \end{array}\!\! \right) \;.
\end{equation} 

\ni Here, $ M_{e \mu} = - M_{e \tau}$, $ M_{\mu \mu} = M_{\tau \tau}$ and 

%rownanie 10
\begin{eqnarray} 
 M_{e e} = c^2_{12}m_1 + s^2_{12} m_2 \;,\; M_{e e} & \!\! + \!\! & M_{\mu \mu} - M_{\mu \tau} = m_1 + m_2 \;,\; M_{\mu \mu} + M_{\mu \tau} = m_3 \;,\nonumber \\  M_{e \mu} & \!\! = \!\! & (m_2 - m_1)c_{12} s_{12}/\sqrt{2}\;.
\end{eqnarray}

Assuming that $ M_{e e} = 0 $, we get from Eq. (10) the relations $ M_{\mu  \mu} = (m_3 + m_2 + m_1)/2 $, $ M_{\mu  \tau} = (m_3 - m_2 - m_1)/2 $, $ M_{e  \mu} = (s_{12}/c_{12}) m_2/ \sqrt{2} $, and

%rownanie 11
\begin{equation} 
\frac{m_1}{m_2} = - \frac{s^2_{12}}{c^2_{12}}\; , \; \Delta m^2_{21} \equiv m^2_2 - m^2_1 = m^2_2 \frac{c^2_{12} - s^2_{12}}{c^4_{12}}
\end{equation}

\ni or

%rownanie 12
\begin{equation} 
m_1 = - \sqrt{\Delta m^2_{21}} \frac{s^2_{12}}{\sqrt{c^2_{12} - s^2_{12}}}\; , \; 
m_2 =  \sqrt{\Delta m^2_{21}} \frac{c^2_{12}}{\sqrt{c^2_{12} - s^2_{12}}}\; , 
\end{equation}

\ni when taking $ m_1 \leq m_2 $. For instance, applying to Eq. (12) the LOW solar solution [4] {\it i.e.}, $ s^2_{12} \sim 0.5 - 0.15 $, $ c^2_{12} \sim 0.5 + 0.15 $ and $ \Delta m^2_{21} \sim 1.0\times 10^{-7}\,{\rm eV}^2 $, we estimate

%rownanie 13
\begin{equation} 
m_1  \sim - 2.0 \times 10^{-4}\,{\rm eV}\; , \; m_2 \sim 3.8 \times 10^{-4}\,{\rm eV}\;  , 
\end{equation}

\ni while the Super--Kamiokande result $ \Delta m^2_{32} \sim 3.5 \times 10^{-3}\,{\rm eV}^2 $ [5] leads to the estimation

%rownanie 14
\begin{equation} 
m_3  \sim 5.9 \times 10^{-2}\,{\rm eV}\; , 
\end{equation}

\ni what shows explicitly that $ | m_1| \stackrel{<}{\sim} m_2 \ll m_3 $. Thus,    in this case

\vspace{-0.2cm}

%rownanie 15
\begin{equation} 
M_{e e} =0 , M_{\mu \mu} = M_{\tau \tau}  \sim 3.0 \times 10^{-2}{\rm eV} , M_{e \mu} =  -M_{e \tau} \sim 1.9 \times 10^{-4}{\rm eV}, M_{\mu \tau} \sim 3.0 \times 10^{-2}{\rm eV},
\end{equation}

\ni where $  M_{\mu \mu} \stackrel{>}{\sim} M_{\mu \tau} \gg M_{e \mu}$.

In conclusion, the two--mixing texture of three (Dirac or Majorana) active neutrinos $ \nu_\alpha \;\,(\alpha = e\,,\,\mu\,,\,\tau) $, described by the formulae (2) and (5), is neatly consistent with the observed solar and atmospheric neutrino deficits, but it predicts no LSND effect whose confirmation should imply, therefore, the existence of at least one sterile neutrino $ \nu_s $, mixing with $ \nu_e $. This might be either one extra,  light (Dirac or Majorana) sterile neutrino $ \nu_s $ [6,10] or one of three conventional, light Majorana sterile neutrinos $ \nu_\alpha^{(s)} = \nu_{\alpha R} + (\nu_{\alpha R})^c\;\;(\alpha = e\,,\,\mu\,,\,\tau)$ [11,12] existing in this case beside three light Majorana active neutrinos $\nu_\alpha^{(a)} = \nu_{\alpha L} + (\nu_{\alpha L})^c \;\;(\alpha = e\,,\,\mu\,,\,\tau)$ [of course, $\nu_\alpha^{(a)} = \nu_{\alpha L}$ and $ \nu_{\alpha L}^{(s)} = (\nu_{\alpha R})^c] $.

The essential agreement of the observed neutrino oscillations with the two--mixing option (2) for $ U $ (provided there is really no LSND effect) suggests that the conjecture of absence of direct mixing of massive neutrinos $ \nu_1 $ and $ \nu_3 $, leading to $ U $ of the form (2), is somehow physically important. This absence tells us that {\it only the close neighbours, $ \nu_1$ and $\nu_2$, $\nu_2$ and $\nu_3 $, in the hierarchy of massive neutrinos $ \nu_1\,, \,\nu_2\,, \,\nu_3 $ mix directly}.

\vspace{0.2cm}

\ni {\bf 3. The second option}

\vspace{0.2cm}

When we want to introduce one sterile neutrino mixing with three active neutrinos $ \nu_e\,, \,\nu_\mu\,, \,\nu_\tau $ (thus leading to four massive neutrino states $ \nu_0\,, \, \nu_1\,, \,\nu_2\,, \,\nu_3 $), we ought to extend properly the two--mixing formula (2) of the previous $ 3 \times 3 $ mixing matrix $ U $. A natural form of such a new $4 \times 4 $ mixing matrix seems to be

%rownanie 16
\begin{eqnarray} 
U & = & \left( \begin{array}{cccc} \; c_{01} & s_{01} & 0  & 0  \\ -s_{01} & \; c_{01} & 0 & 0 \\ 0 & 0 & \; c_{23} & s_{23}\\  0 & 0 &  -s_{23} & c_{23} \end{array} \right)  \left( \begin{array}{cccc}1 & 0 & 0 & 0\\ 0 & \;c_{12} & s_{12}  & 0  \\ 0 &-s_{12} & c_{12} & 0 \\ 0 & 0 & 0 & 1 \end{array} \right) = 
 \left( \begin{array}{rrrr} c_{01} & s_{01}c_{12}  & s_{01}s_{12}  & 0\;\, \\ -s_{01} & c_{01} c_{12}  & c_{01}s_{12}  & 0\;\, \\ 0\;\, & -c_{23}s_{12} & c_{23} c_{12} & s_{23} \\ 0\;\, & s_{23}s_{12}  & -s_{23} c_{12} & c_{23} \end{array} \right) \,, \nonumber \\ & &
\end{eqnarray} 

\ni if in the hierarchy of massive neutrinos $ \nu_0\,, \, \nu_1\,, \,\nu_2 \,, \, \nu_3 $ the new massive neutrino $ \nu_0 $ mixes directly only with its close neighbour $ \nu_1 $ ($ c_{01} = \cos \theta_{01} $ and $ s_{01} = \sin \theta_{01}$). Then, {\it only the close neighbours, $\nu_0 $ and $\nu_1 $, $\nu_1$ and $\nu_2 $, $\nu_2$ and $\nu_3$, in the hierarchy of massive neutrinos $\nu_0\,, \,\nu_1\,, \,\nu_2\,, \,\nu_3$ mix directly}.  In the limiting case of $\theta_{01} = 0 $ the three--mixing form (16) of $ 4\times 4 $ mixing matrix is reduced to the two--mixing form (2) of $ 3\times 3 $ mixing matrix. It is interesting to observe that in this four--neutrino texture the sterile small--angle MSW global solution [4] leads to a small value $\theta_{01} \simeq s_{01} \sim 0.0017 $ ({\it cf.} the first relation (19) later on). If, however, a considerable or even nearly maximal mixing of  $\nu_0 $ and $\nu_1 $ can work effectively for solar  $\nu_e $'s, such a small value of $\theta_{01} $ may be replaced by a considerable $\theta_{01} $ or even $\theta_{01} \simeq \pi /4 $: $ c_{01} \stackrel{>}{\sim} 1/\sqrt{2} \stackrel{>}{\sim}  s_{01} $. On the other hand, a small mixing of $\nu_1$ and $\nu_2 $ may be sufficient to explain the possible LSND effect (or its modified version), while the nearly maximal mixing of $\nu_2$ and $\nu_3 $ still works well for atmospheric $\nu_\mu $'s. Thus, putting in Eq. (16) $ c_{23} \simeq 1/\sqrt{2} \simeq s_{23}$ and $ c_{12} \simeq 1 \gg s_{12} \simeq \varepsilon \sqrt{2} > 0 $, we get approximately from Eq. (16)

%rownanie 17
\begin{equation} 
U = \left( \begin{array}{cccc} \; c_{01} & s_{01} & \varepsilon  & 0  \\ -s_{01} & \; c_{01} & \varepsilon & 0 \\ 0 & \varepsilon & \; 1/\sqrt{2} & 1/\sqrt{2} \\  0 & \varepsilon  &  -1/\sqrt{2}  & 1/\sqrt{2} \end{array} \right) \;.
\end{equation}

The mixing matrix (17) gives through Eqs. (6) with (7), where now $ \alpha, \beta = s\,,\, e\,,\,\mu\,,\, \tau $ and $ i,j = 0,1,2,3 $, the following neutrino oscillation probabilities:

%rownanie 18
\begin{eqnarray}
P(\nu_e \rightarrow \nu_e) & = & 1 - (2 c_{01} s_{01})^2  \sin^2 x_{10} - 4 \varepsilon^2(s^2_{01} \sin^2 x_{21} +c^2_{01} \sin^2 x_{31}) \nonumber \\
& \simeq & 1 - (2 c_{01} s_{01})^2  \sin^2 x_{10} - 2 \varepsilon^2 \;,
\nonumber \\ 
P( \nu_\mu \rightarrow \nu_\mu) & = & 1 - \sin^2 x_{23} - 2\varepsilon^2 (\sin^2 x_{21} + \sin^2 x_{31}) \simeq 1 - \sin^2 x_{23} - 2 \varepsilon^2    \nonumber \\ 
P(\nu_\mu \rightarrow \nu_e) & = &  2\sqrt{2}\, c_{01}\varepsilon^2 \sin^2 x_{21} \;.
\end{eqnarray}

\ni The second step in the first and second formula (18) is valid for $ x_{10} = x_{\rm sol} = O(1)$ and $ x_{32} = x_{\rm atm} = O(1)$, respectively, where now in this four--neutrino texture $ m^2_0  \stackrel{<}{\sim} m^2_1 \ll m^2_2 \stackrel{<}{\sim} m^2_3 $ or equivalently $\Delta m^2_{10} \ll \Delta m^2_{21} \stackrel{<}{\sim} \Delta m^2_{20}$ and $\Delta m^2_{32} \ll \Delta m^2_{21} \stackrel{<}{\sim} \Delta m^2_{31}$. 

Then, the formulae (18) are consistent with experimental data for solar $\nu_e $'s [4], atmospheric $\nu_\mu $'s [5] and LSND accelerator $\nu_\mu $'s [1], if 

%rownanie 19
\begin{eqnarray}
(2 c_{01} s_{01})^2  \leftrightarrow   \sin^2 2\theta_{\rm sol} \sim \left\{
\begin{array}{l} 6.6\times 10^{-3}\;{\rm or} \\ 0.72 \;{\rm or} \\ 0.90 \end{array} \right. & \!,\! & \Delta m^2_{10} \leftrightarrow \Delta m^2_{\rm sol} \sim \left\{
\begin{array}{l} 4.0 \times 10^{-6}\;\;{\rm eV}^2  \;{\rm or}\\ 6.5\times 10^{-11} \;{\rm eV}^2  \;{\rm or} \\ 4.4\times 10^{-10}\;{\rm eV}^2 \end{array} \right.\;\; , \nonumber \\
1  \leftrightarrow \sin^2 2\theta_{\rm atm} \sim 1\;\;\;\;\;\;\;\;\;\;\;\;\;\;\;\;\;\;\;\;\;\;\, & \!,\! & \Delta m^2_{32}   \leftrightarrow \Delta m^2_{\rm atm}  \sim 3.5 \times 10^{-3}\;{\rm eV}^2 \;,
\nonumber \\ 
2\sqrt{2} \varepsilon^2 \leftrightarrow  \sin^2 2\theta_{\rm LSND} \sim 10^{-2} \;\;\;\;\;\;\;\;\;\;\;\;\;\;\; & \!,\! & \Delta m^2_{21} \leftrightarrow \Delta m^2_{\rm LSND} \sim 1\;{\rm eV}^2 \;,
\end{eqnarray}

\ni respectively. Here, in the case of solar $\nu_e$'s we apply the sterile small--angle MSW global solution or, just for an illustration, smaller--mass or larger--mass vacuum global solution [4] (however, in the case of solar $\nu_e$'s the role of $\nu_s$'s in the disappearance of $\nu_e$'s is recently disputed [4,7]). Then, $c^2_{01} \sim$ (1 or 0.76 or 0.66) and $s^2_{01} \sim$ (0.0017 or 0.24 or 0.34). From Eqs. (19) we obtain readily the estimations $m_2 \sim 1$ eV, $m_3 \sim 1$ eV and $m_1 \sim (2.0 \times 10^{-3}$ or $ 8.1\times 10^{-6}$ or $ 2.1\times 10^{-5}$) eV, the last if we conjecture that $ m_0 = 0 $, and $\varepsilon \sim 5.9\times 10^{-2}$. This shows explicitly that $ m_0  \stackrel{<}{\sim} m_1 \ll m_2 \stackrel{<}{\sim} m_3 $.

In the case of mixing matrix (17), the $4\times 4$ mass matrix $ M = (M_{\alpha \beta})\;\; (\alpha, \beta = s\,,\, e\,,\, \mu\,,\, \tau) $, takes, up to $ O(\varepsilon^2) $, the form

%rownanie 20
\begin{eqnarray} 
\lefteqn{M = } 
& & \;\left(\!\!\begin{array}{cccc}c_{01}^2 m_1\! +\! s^2_{01} m_2  & c_{01} s_{01}(m_1 \!-\! m_0) & \varepsilon(s_{01} m_1 \!+\! m_2/\sqrt{2}) & \varepsilon(s_{01} m_1 \!-\! m_2/\sqrt{2})  \\  c_{01} s_{01}(m_1 \!-\! m_0)
& s^2_{01}m_0 + c^2_{01}m_1 & \varepsilon(c_{01} m_1 \!+\! m_2/\sqrt{2}) & \varepsilon(c_{01} m_1 \!-\! m_2/\sqrt{2})  \\  \varepsilon(s_{01} (m_1 \!+\! m_2/\sqrt{2}) & \varepsilon(c_{01} (m_1 \!+\! m_2/\sqrt{2}) & (m_2 \!+\! m_3) /2 & (m_2 \!-\! m_3) /2 \\  \varepsilon(s_{01} (m_1 \!-\! m_2/\sqrt{2}) & \varepsilon(c_{01} (m_1 \!-\! m_2/\sqrt{2}) & (m_2 \!-\! m_3) /2 & (m_2 \!+\! m_3) /2 \end{array} \right)  \;, \nonumber \\ & &
\end{eqnarray} 

\ni since $M = \left( \sum_i U_{\alpha i} U^*_{\beta i} m_i \right)$. Hence, up to $ O(\varepsilon^2)$,

%rownanie 21
\begin{equation} 
m_{0,1} = \frac{M_{ss} + M_{ee}}{2} \mp \sqrt {\left(\frac{M_{ee} - M_{ss}}{2}  \right)^2 + M^2_{se}}\,,\, m_{2,3} = M_{\mu \mu} \mp M_{\mu \tau}\,,
\end{equation} 

\ni where $ M_{ee} = M_{ss} + (c_{01}/s_{01} - s_{01}/c_{01})M_{se}$, $  M_{ss} = ( s_{01}/c_{01})M_{se}$ (when $m_0 = 0$), $ M_{se} \sim  c_{01} s_{01}(2.0 \times 10^{-3} $ or $8.1\times 10^{-6}$ or $2.1\times 10^{-5})$ eV (when $m_0 = 0$) and $ M_{\tau \tau} = M_{\mu \mu} \sim 1$ eV, $ M_{\mu \tau} \sim (3.5/4)\times 10^{-3}$ eV.

If eventually the LSND effect turns out to be confirmed, then at least one sterile neutrino mixing with three active neutrinos ought to exist. The second option discussed here is a natural candidate for its texture. If there are more sterile neutrinos mixing with active neutrinos, the neutrino texture would be effectively more extended [13].

\vfill\eject

~~~~
\vspace{0.5cm}

{\centerline{\bf References}}

\vspace{0.5cm}

{\everypar={\hangindent=0.5truecm}
\parindent=0pt\frenchspacing

{\everypar={\hangindent=0.5truecm}
\parindent=0pt\frenchspacing

~1.~C.~Athanassopoulos {\it et al.} (LSND Collaboration), {\it Phys. Rev. Lett.} {\bf 75}, 2650 (1995); {\it Phys. Rev.} {\bf C 54}, 2685 (1996); {\it Phys. Rev. Lett.} {\bf 77}, 3082 (1996); {\bf 81}, 1774 (1998); G. Mills, Talk at {\it Neutrino 2000}, Sudbury, Canada, June 2000.

\vspace{0.2cm}

~2.~{\it Cf. e.g.} F. Feruglio, {\it Acta Phys. Pol.} {\bf B 31}, 1221 (2000); and references therein.

\vspace{0.2cm}

~3.~W. Kr\'{o}likowski, hep--ph/0007255.

\vspace{0.2cm}

~4.~{\it Cf. e.g.}~J.N.~Bahcall, P.I.~Krastev and A.Y.~Smirnov, {\it Phys. Lett.} {\bf B 477}, 401 (2000); hep--ph/0002293.

\vspace{0.2cm}

~5.~Y.~Fukuda {\it et al.} (Super--Kamiokande Collaboration), {\it Phys. Rev. Lett.} {\bf 81}, 1562 (1998) [E. {\bf 81}, 4279 (1998)]; {\bf 82}, 1810 (1999); {\bf 82}, 2430 (1999).

\vspace{0.2cm}

~6.~{\it Cf. e.g.}~G. Giunti, M.C. Gonzales--Garcia and C. Pe\~{n}a--Garay,  hep--ph/0001101; and references therein.

\vspace{0.2cm}

~7.~{\it Cf. e.g.} W. Barger, {\it et al.}, hep--ph/0008019.

\vspace{0.2cm}

~8.~Z. Maki, M. Nakagawa and S. Sakata, {\it Progr. Theor. Phys.} {\bf 28}, 870 (1962).

\vspace{0.2cm}

~9.~M. Appolonio {\it et al.} (Chooz Collaboration), {\it Phys. Lett.} {\bf B 420}, 397 (1998); {\bf B 466}, 415 (1999).

\vspace{0.2cm}

10.~{\it Cf. e.g.} W. Kr\'{o}likowski, {\it Nuovo Cim.} {\bf A 111}, 1257 (1999);  and references therein.

\vspace{0.2cm}

11.~{\it Cf. e.g.} A. Geiser, hep--ph/9901433; and references therein. 

\vspace{0.2cm}

12. {\it Cf. e.g.} W. Kr\'{o}likowski, {\it Nuovo Cim.} {\bf A 112}, 893 (1999), also hep--ph/9904489; {\it Acta Phys. Pol.} {\bf B 31}, 663 (2000);  and references therein.

\vspace{0.2cm}

13.~W. Kr\'{o}likowski, hep--ph/0004222; and references therein.

%\vspace{0.2cm}

\vfill\eject

\end{document}